\documentclass[10pt,a4paper]{article}

\usepackage{epsf,graphicx,amssymb,amsfonts,amsmath,hyperref,fullpage}

\begin{document}

\title{Power Divergences in Overlapping Wilson Lines}

\author{Matthias Berwein\\
Physik-Department, Technische Universit\"{a}t M\"{u}nchen\\James-Franck-Str.~1, 85748 Garching, Germany}

\maketitle

\begin{abstract}
 We discuss the divergence structure of Wilson line operators with partially overlapping segments on the basis of the cyclic Wilson loop as an explicit example. The generalized exponentiation theorem is used to show the exponentiation and factorization of power divergences for certain linear combinations of associated loop functions.
\end{abstract}

\section{Introduction}
Wilson line operators and correlators thereof have a wide range of applications, one of which is in the context of asymptotically static particles: the propagator of a static particle interacting with a gauge field is given essentially by a temporal Wilson line.
\begin{equation}
 D(x_1,x_2)=\theta(t_1-t_2)\delta^{(3)}(\vec{x}_1-\vec{x}_2)\mathcal{P}\exp\left[ig\int_{t_2}^{t_1}dt\,A_0(t,\vec{x}_1)\right]\,,
\end{equation}
where $\mathcal{P}$ stands for path ordering. Another useful property of Wilson lines is that they transform under gauge transformations as $W(x_1,x_2)\to V(x_1)W(x_1,x_2)V^\dagger(x_2)$, which makes it possible to construct gauge invariant operators out of spatially separated fields.

It follows from these properties that a rectangular Wilson loop is related to the potential between two static sources and therefore it has been studied in great detail in the context of heavy quarkonium~\cite{RWL1,RWL2,RWL3,RWL4,RWL5,RWL6}. It is known that rectangular Wilson loops are divergent quantities: there appear logarithmic divergences that can be removed through charge renormalization, other logarithmic divergences related to the appearance of cusps in the contour and power divergences proportional to the length of the contour. The latter two can be removed through a multiplicative constant~\cite{Ren1,CuspDiv}.

The free energy of a static quark-antiquark pair in a quark-gluon-plasma is related to a different Wilson line operator~\cite{PLC1,PLC2}, the Polyakov loop correlator:
\begin{equation}
 P_c(r,T)=\frac{1}{N_c^2}\left\langle\mathrm{Tr}[P(\vec{r},T)]\mathrm{Tr}[P^\dagger(\vec{0},T)]\right\rangle\,,\hspace{10pt}\mathrm{with}\hspace{10pt}P(\vec{r},T)=\mathcal{P}\exp\left[ig\int_{0}^{1/T}d\tau\,A_0(\tau,\vec{r})\right]\,.
\end{equation}
We have introduced the Polyakov loop operator $P(\vec{r},T)$, the trace of which is gauge invariant because of the periodic boundary conditions of the imaginary time formalism. The free energy is then given by $-T\log P_c(r,T)$ and it contains all possible colour configurations of the quark and the antiquark. This is why it has sometimes been called the ``colour-averaged potential''. Since the combination of a colour triplet and an antitriplet can be decomposed into a singlet and an octet, one can also define singlet and octet free energies, however, there are different possibilities to do so. One of them is to define the singlet free energy through the logarithm of $\left\langle\mathrm{Tr}[P(\vec{r},T)P^\dagger(\vec{0},T)]\right\rangle/N_c$ calculated in Coulomb gauge. This definition depends on the choice of Coulomb gauge, a gauge-independent alternative is to introduce spatial Wilson lines $S(\vec{r})=\mathcal{P}\exp\left[ig\int_{1}^{0}ds\,\vec{r}\cdot\vec{A}(0,s\vec{r})\right]$ to connect the Polyakov loops. This is called the cyclic Wilson loop:
\begin{equation}
 W_c(r,T)=\frac{1}{N_c}\left\langle\mathrm{Tr}[P(\vec{r},T)S(\vec{r})P^\dagger(\vec{0},T)S^\dagger(\vec{r})]\right\rangle\,.
\end{equation}

The Coulomb correlator contains quark-antiquark states that transform as a singlet under global but not local gauge transformations and is comparatively simple to calculate. The choice of Coulomb gauge ensures that the associated singlet free energy approaches the vacuum static energy for $rT\ll1$ and removes any logarithmic divergences not related to charge renormalization. The cyclic Wilson loop gets rid of the gauge dependence, but the introduced Wilson lines $S(\vec{r})$ give new contributions which include logarithmic divergences that cannot be removed by charge renormalization or a multiplicative constant~\cite{Laine}. It was shown in~\cite{RCWL} that in fact the difference $W_c-P_c$ is a multiplicatively renormalizable quantity, but this treatment was based on dimensional regularization and therefore ignored power divergences. A treatment of these can be found in~\cite{Loops}, where it is shown that the multiplicative renormalizability of $W_c-P_c$ does not rely on the regularization scheme. In these proceedings we will summarize the arguments which lead to that result and comment on the general case of overlapping Wilson lines.

\section{Divergences in the cyclic Wilson loop}

The key difference between the cyclic Wilson loop and other rectangular Wilson loops is that because of the periodic boundary conditions of the imaginary time formalism the spatial lines of the cyclic loop overlap exactly. Therefore it has intersections instead of cusps. It is known that intersections in Wilson loops lead to a different kind of divergence than cusps, and in order to remove that divergence one has to introduce mixing between different associated loops~\cite{Ren2}. Associated loops are given by the same Wilson lines that are connected differently at the intersection.

One can treat the cyclic Wilson loop as having only two intersections where the strings $S$ connect to the Polyakov loops $P$. Points along the strings are in principle also intersections, but they only contribute trivially to the mixing and are therefore irrelevant. The associated loops obtained from the cyclic Wilson loop are given by all possibilities to connect the strings and Polykov loops into closed loops, and all turn out to be equal to the Polykov loop correlator:
\begin{equation}
 \frac{1}{N_c^2}\left\langle\mathrm{Tr}[P]\mathrm{Tr}[SP^\dagger S^\dagger]\right\rangle=\frac{1}{N_c^2}\left\langle\mathrm{Tr}[S^\dagger PS]\mathrm{Tr}[P^\dagger]\right\rangle=\frac{1}{N_c^3}\left\langle\mathrm{Tr}[P]\mathrm{Tr}[S^\dagger S]\mathrm{Tr}[P^\dagger]\right\rangle=P_c\,.
\end{equation}
So though in principle one has to mix four loop functions, this particular case reduces to a mixing between the cyclic Wilson loop and the Polykov loop correlator:
\begin{equation}
 \left(\begin{array}{c} W_c^{(R)} \\ P_c^{(R)} \end{array}\right)=\left(\begin{array}{cc} Z_{int} & 1-Z_{int} \\ 0 & 1 \end{array}\right)\left(\begin{array}{c} W_c \\ P_c \end{array}\right)\,,
\end{equation}
where here $W_c^{(R)}$ and $P_c^{(R)}$ denote the partially renormalized quantities where only the intersection divergences have been subtracted. The exact form of the renormalization matrix, in particular the fact that it only depends on one renormalization constant $Z_{int}$, follows from considerations that in the initial ansatz with four loop functions the $4\times4$ renormalization matrix has to be the tensor product of two identical $2\times2$ matrices from each intersection, from the convention that at leading order in $\alpha_s$ the renormalization matrix should be the unit matrix, and from the fact that the Polyakov loop correlator does not have any intersection or cusp divergences.

By diagonalizing the renormalization matrix we see that the difference $W_c-P_c$ is free of intersection divergences after multiplying with a suitable $Z_{int}$. In the $\overline{\mathrm{MS}}$-scheme that constant is given by
\begin{equation}
 Z_{int}^{\overline{\mathrm{MS}}}=\exp\left[-\frac{N_c\alpha_s}{\pi\bar{\varepsilon}}+\mathcal{O}(\alpha_s^2)\right]\,.
\end{equation}
Since this scheme is based on dimensional regularization, where power divergences do not appear, $Z_{int}^{\overline{\mathrm{MS}}}(W_c-P_c)$ is completely free of UV divergences.

In other regularization schemes power divergences do appear and for non-overlapping Wilson lines they take the form of an exponential of some linearly diverging constant $\Lambda_R$ times the length of the contour~\cite{Smooth}. $\Lambda_R$ may depend on the representation $R$ and the regularization scheme, but not on any other property of the Wilson line. The Polyakov loop correlator is therefore free of power divergences after multiplication with $\exp\left[-2\Lambda_F/T\right]$. If the cyclic Wilson loop had no overlapping sides, its power divergences would be removed accordingly by $\exp\left[-2\Lambda_F/T-2\Lambda_Fr\right]$, but just like the cusp divergences were turned into intersection divergences by the periodic boundary conditions, also the structure of the power divergences is changed in such a way that they factorize not for $W_c$ alone but again for the difference $W_c-P_c$. This can be shown by reexpressing the difference of the two loop functions $W_c$ and $P_c$ by a single loop function.

The product of the two strings $S(\vec{r})$ and $S^\dagger(\vec{r})$ can be decomposed in the same way that the product of a fundamental and an antifundamental representation can be decomposed into a singlet and an adjoint representation.
\begin{equation}
 S^\dagger_{li}(\vec{r})S^{\phantom{\dagger}}_{jk}(\vec{r})=S^\dagger_{li'}(\vec{r})\left(\frac{\delta_{ji}\,\delta_{i'j'}}{N_c}+\frac{T^a_{ji}\,T^a_{i'j'}}{T_F}\right)S_{j'k}(\vec{r})=\frac{\delta_{ji}\,\delta_{lk}}{N_c}+\frac{T^a_{ji}\,S_A^{ab}(\vec{r})\,T^b_{lk}}{T_F}\,,
\end{equation}
where $S_A(\vec{r})$ is a spatial Wilson line in the adjoint representation and the colour matrices are normalized such that $\mathrm{Tr}\bigl[T^aT^b\bigr]=T_F\delta^{ab}$. The first term in the brackets is the projector on the singlet representation, the second term is the projector on the adjoint representation, and the sum of both gives the identity $\delta_{i'i}\delta_{jj'}$. If we insert this result into the definition of the cyclic Wilson loop, then the first part gives exactly the Polyakov loop correlator, so we get for the difference
\begin{equation}
 W_c-P_c=\frac{1}{N_c}\left\langle P^{\phantom{\dagger}}_{ij}\,S^{\phantom{\dagger}}_{jk}\,P^\dagger_{kl}\,S^\dagger_{li}\right\rangle-\frac{1}{N_c^2}\left\langle P^{\phantom{\dagger}}_{ii}\,P^\dagger_{kk}\right\rangle=\frac{T^a_{ji}\,T^b_{lk}}{N_cT_F}\left\langle P^{\phantom{\dagger}}_{ij}\,S^{ab}_A\,P^\dagger_{kl}\right\rangle\,.
 \label{comps}
\end{equation}

The last line now contains the expectation value of three untraced Wilson lines. The advantage of writing the result in this way is that there exists an exponentiation formula for untraced Wilson lines in general colour representations. It has been known for a long time~\cite{Exp1,Exp2} that expectation values of a closed Wilson line can be exponentiated, i.e.~they can be expressed as an exponential of a series of Feynman diagrams. These diagrams are the same as those that would appear in a straightforward perturbative calculation of the loop, but there appear less of them in the exponent and they have different coefficients. Recently, this exponentiation property has been generalized in~\cite{Exp3,Exp4}.

This generalized exponentiation is to be understood in the following way. Feynman diagrams for untraced Wilson lines have a number of initial and final colour indices, where \textit{initial} and \textit{final} are used in the context of the path ordering of the Wilson lines. For instance, in the last part of eq.~\eqref{comps} $i$, $a$ and $k$ are final indices and $j$, $b$ and $l$ are initial indices. A multiplication of two diagrams can be defined as the contraction of the initial indices of the first diagram with the final indices of the second, and an exponential of diagrams is then defined in the usual way as a power series with respect to this multiplication.

This tensor exponentiation can be put in the form of a standard matrix exponentiation, but for the present line of argument it is not necessary calculate this exponential explicitly. The unit element of this multiplication is the tensor $\delta_{ij}\delta^{ab}\delta_{kl}$. The diagrams that appear in the exponentiated expression of the untraced Wilson lines in eq.~\eqref{comps} can be split into a part that is proportional to this unit tensor and a part that is not. It is then straightforward to see that the coefficient of the unit tensor factorizes out of the exponential:
\begin{equation}
 \left\langle P_{ij}(\vec{r},T)S^{ab}_A(\vec{r})P^\dagger_{kl}(\vec{0},T)\right\rangle=\exp\left[X(r,T)\delta_{ij}\delta^{ab}\delta_{kl}+Y_{ij,kl}^{ab}(r,T)\right]=\exp[X(r,T)]\exp\left[Y_{ij,kl}^{ab}(r,T)\right]\,,
\end{equation}
where in the last term the first exponential is to be understood as the usual exponential function of the number $X$ and the second as the tensor exponential defined above. It then remains to show that all power divergences that appear in the exponent are proportional to the unit tensor.

In the case of the exponentiation of a closed Wilson loop there is a simple criterion to determine which diagrams can appear in the exponent: if one can cut the Wilson loop at two points in such a way that each of the resulting pieces couples to gluons, but there are no gluons that connect the two pieces to each other, then this diagram does not appear in the exponent~\cite{Exp1,Exp2}. For the exponentiation of untraced Wilson lines there exists a similar criterion: if any set of gluons can be ``cut out'' of a Wilson line, in the sense that one can cut the same Wilson line once or twice such that each of the resulting pieces couples to gluons, but at least one of those pieces is not connected to any other piece or any of the other Wilson lines by gluons, then this diagram does not contribute to the exponent. The same is true if at least two Wilson lines couple to gluons but are not connected to each other~\cite{Exp3,Exp4}.

It has been shown in~\cite{Ren1} that power divergences come from sets of gluons that can be contracted to a point on the Wilson line. Such a set of gluons does not connect to another Wilson line, so that diagram can only contribute to the exponent if all gluons connect to the same Wilson line. If the set of gluons can be cut out of that line, then the diagram also does not contribute. So the only source of power divergences in the exponent are diagrams where all gluons attach to the same Wilson in such a way that none can be cut out. These diagrams are always proportional to the unit tensor, because a Wilson line without gluons attached to it just gives just a Kronecker delta, and the expectation value of a single Wilson line is always proportional to the unit matrix in any representation~\cite{Ren1}.

So the power divergences factor out of the exponential and this happens individually for each Wilson line, with each factor given by the exponential of a linearly divergent constant $\Lambda_R$ depending on the representation times the length of the respective Wilson line. This shows that the difference $W_c-P_c$ can be completely renormalized by a multiplicative constant
\begin{equation}
 W_c^{(R)}-P_c^{(R)}=\exp\left[-2\Lambda_F/T-\Lambda_Ar\right]Z_{int}(W_c-P_c)\,,
\end{equation}
\begin{equation}
 P_c^{(R)}=\exp\left[-2\Lambda_F/T\right]P_c\,.
\end{equation}
These expressions are in particular also valid for lattice evaluations. So while in the case of the Polyakov loop correlator it is sufficient to consider the ratio of $P_c(r,T)/P_c(r_0,T)$ at the same temperature $T$ with some arbitrary reference distance $r_0$ to remove all divergences, one has to keep in mind that for the cyclic Wilson loop this procedure does not work. First of all one should work with the difference $W_c-P_c$ so that the divergences factorize, but even then a ratio of different values for $r$ is not free of divergences due to the factor $\exp[\Lambda_Ar]$, so one has to find other ways to remove that divergence.

The previous arguments can be extended without problems to the general case of Wilson line operators with overlapping segments. The overlapping Wilson lines can be decomposed into a sum of Wilson lines in different colour representations. For each of these terms the linear divergences factorize in an exponential of a representation-dependent constant times the length of the overlapping segment. In general there will also be intersection divergences at the endpoints of the overlapping segments, so one has to mix associated Wilson line operators. Each of these associated operators will have a different decomposition of the overlapping segments, so one can construct linear combinations of the associated operators that include only one representation. It remains to be seen, whether also the renormalization matrix for the intersection divergences will always be diagonal in this basis, or if this was a coincidence for $W_c$ and $P_c$.

\section*{Acknowledgments}
This work was supported by the DFG grant Br 4058/1-1 and the DFG cluster of excellence \textit{Origin and Structure of the Universe} (www.universe-cluster.de).

\end{document}